\documentstyle[12pt]{article}

        {\newpage
        \setcounter{page}{1}}

\renewcommand{\title}[1]{%
        {\begin{center}
        \Large\bf #1
        \end{center}}
        \vskip .3in}

\renewcommand{\author}[1]{%
        {\begin{center}
        #1
        \end{center}}}

\renewcommand{\abstract}[1]{%
        \begin{center}%
        {\vspace{1em}\vspace{0pt}\bf Abstract}%
        \end{center}%
        \noindent #1}

\renewcommand{\date}[1]{%
        \begin{center}%
        #1%
        \end{center}}

        {\end{thebibliography}}

\makeatletter
        \@addtoreset{equation}{section}%
\makeatother

\newcommand{\beq}{\begin{eqnarray}}
\newcommand{\eeq}{\end{eqnarray}}

\newcommand{\mybar}[1]%
        {\kern 0.8pt\overline{\kern -0.8pt#1\kern -0.8pt}\kern 0.8pt}
\newcommand{\sla}[1]%
        {\raise.15ex\hbox{$/$}\kern-.57em #1}
\newcommand{\roughly}[1]%
        {\mathrel{\raise.3ex\hbox{$#1$\kern-.75em\lower1ex\hbox{$\sim$}}}}

\newcommand{\drawsquare}[2]{\hbox{%
\rule{#2pt}{#1pt}\hskip-#2pt
\rule{#1pt}{#2pt}\hskip-#1pt
\rule[#1pt]{#1pt}{#2pt}}\rule[#1pt]{#2pt}{#2pt}\hskip-#2pt
\rule{#2pt}{#1pt}}

\newcommand{\Yfund}{\raisebox{-.5pt}{\drawsquare{6.5}{0.4}}}
\newcommand{\Ysymm}{\raisebox{-.5pt}{\drawsquare{6.5}{0.4}}\hskip-0.4pt%
        \raisebox{-.5pt}{\drawsquare{6.5}{0.4}}}
\newcommand{\Yasymm}{\raisebox{-3.5pt}{\drawsquare{6.5}{0.4}}\hskip-6.9pt%
        \raisebox{3pt}{\drawsquare{6.5}{0.4}}}
\newcommand{\Ythreea}{\raisebox{-3.5pt}{\drawsquare{6.5}{0.4}}\hskip-6.9pt%
        \raisebox{3pt}{\drawsquare{6.5}{0.4}}\hskip-6.9pt
        \raisebox{9.5pt}{\drawsquare{6.5}{0.4}}}

\newcommand{\jref}[4]{{\it #1} {\bf #2}, #3 (#4)}

\newcommand{\MPLA}[3]{\jref{Mod.\ Phys.\ Lett.}{A#1}{#2}{#3}}

\newcommand{\NPB}[3]{\jref{Nucl.\ Phys.}{B#1}{#2}{#3}}

\newcommand{\PLB}[3]{\jref{Phys.\ Lett.}{#1B}{#2}{#3}}

\newcommand{\PRD}[3]{\jref{Phys.\ Rev.}{D#1}{#2}{#3}}

\begin{document}

\begin{titlepage}
\begin{center}
{\hbox to\hsize{hep-th/9610139 \hfill  MIT-CTP-2581}}
{\hbox to\hsize{               \hfill  BUHEP-96-41}}

\bigskip
\bigskip
\bigskip
             
{\Large \bf  A Systematic Approach to Confinement 
                     in $N=1$ Supersymmetric Gauge Theories} \\

\bigskip
\bigskip
\bigskip

{\bf Csaba Cs\'aki$^a$, Martin Schmaltz$^b$ and  Witold Skiba$^a$}\\

\bigskip

{ \small \it $^a$ Center for Theoretical Physics

Laboratory for Nuclear Science and Department of Physics

Massachusetts Institute of Technology

Cambridge, MA 02139, USA }

{\tt csaki@mit.edu, skiba@mit.edu}
\bigskip 

{\small \it $^b$ Department of Physics

Boston University

Boston, MA 02215, USA 

{\tt schmaltz@abel.bu.edu} }

\vspace{1cm}
{\bf Abstract}\\
\end{center}

\bigskip
\bigskip

We give necessary criteria for $N=1$ supersymmetric theories to be in a
smoothly confining phase without chiral symmetry breaking and with a
dynamically generated superpotential. Using our general arguments we
find all such confining $SU$ and $Sp$ theories with a single gauge group
and no tree level superpotential. 

\bigskip

\end{titlepage}

Following the initial breakthrough in the works of Seiberg on exact
results in $N=1$ supersymmetric QCD (SQCD) \cite{Seib},
much progress has been made in extending these results to other
theories with different gauge and matter fields [2-11].
We now have a whole zoo of examples of supersymmetric theories for
which we know results about the vacuum structure and the infrared
spectrum. A number of theories are known to have dual
descriptions, others are known to confine with or without chiral symmetry
breaking, and some theories do not possess a stable ground state.

Unfortunately, we are still lacking a systematic and general approach
that allows one to determine the infrared properties of a given theory.
The results in the literature have mostly been obtained by an ingenious
guess of the infrared spectrum. This guess is then justified by performing
a number of non-trivial consistency checks which include matching
of the global anomalies, detailed study of the moduli space of vacua, and
the behavior of the theory under perturbations.

In this letter, we will depart from the customary trial and error procedure
and give some general arguments which allow us to classify
a subset of supersymmetric theories. To be specific, we intend to answer
the general question of which supersymmetric field theories may be confining
without chiral symmetry breaking and with a confining superpotential.
We present a few
simple arguments which allow us to rule out most theories
as possible candidates for confinement without chiral symmetry breaking.
For the most part, these arguments already exist in the literature but our
systematic way of putting them to use is new.
As a demonstration of the power of our arguments we give a complete list
of all $SU(N)$ and $Sp(N)$ gauge theories with no tree level superpotential
which confine without chiral symmetry breaking, and we determine the confined
degrees of freedom and the superpotential describing their interactions
(``confining superpotential'').

To begin, let us first explain what we mean by ``smooth confinement without
chiral symmetry breaking and with a non-vanishing confining superpotential",
which, from now on, we will abbreviate by s-confinement.
We will call a theory confining when its infrared physics can be described
exactly in terms of gauge invariant composites and their interactions.
This description has to be valid everywhere on the moduli space of vacua. 
Our definition of s-confinement also requires that the theory dynamically
generates a confining superpotential, which excludes models of the type
presented in Ref.~\cite{ISS}.
Furthermore, the phrase ``without chiral symmetry breaking" implies that the
origin of the classical moduli space is also a vacuum in the quantum theory.
In this vacuum, all the global symmetries of the ultraviolet remain unbroken.
Finally, the confining superpotential is a holomorphic function of the confined
degrees of freedom and couplings, which
describes all the interactions in the extreme infrared.
Note that this definition excludes theories which are in a
Coulomb phase on a submanifold of the moduli space~\cite{phases},
or theories which have distinct Higgs and confining phases with
associated phase boundaries on the moduli space.

Our prototype example for an s-confining theory is Seiberg's SQCD \cite{Seib}
with the number of flavors $F$ chosen to equal $N+1$, where $N$ is the number
of colors, and a ``flavor" is a pair of matter fields in the fundamental and
antifundamental representations of $SU(N)$.
Seiberg argued that the matter
fields $Q$ and $\bar{Q}$ are confined into ``mesons" $M=Q\bar{Q}$ and
``baryons" $B=Q^N$, $\bar{B}=\bar{Q}^N$. At the origin of moduli space all
components of the mesons and baryons are massless and interact via the
confining superpotential
\beq
  W={1 \over \Lambda^{2N-1}}(\det(M)-B M \bar{B}) .
  \label{SQCDpot}
\eeq
At this point, the full global $SU(N+1) \times SU(N+1) \times U(1)_R \times
U(1)$ global symmetry of the model is unbroken, and it is a non-trivial consistency
check that all global anomalies are matched by the mesons and
baryons. The equations of motion $M^{-1} \det(M) -B \bar{B}=0$,
$M \bar{B} = 0$, and $B M = 0$, when expressed in terms of the original
degrees of freedom, $Q$ and $\bar{Q}$, are identical to the classical
constraints. This constitutes another consistency check:
the quantum theory should reproduce these constraints in the classical
limit, $\Lambda \rightarrow 0$, or for generic large vacuum expectation
values (VEVs) which completely break the gauge group.

Other examples in the literature for theories which s-confine
include $SU(N)$ with an antisymmetric tensor, $N-4$ antifundamentals,
and four flavors \cite{sua}, $Sp(2N)$ with $2N+4$ fundamentals
\cite{IntPoul}, $Sp(2N)$ with an antisymmetric tensor and six
fundamentals \cite{Cho,oursp}, a few $SO(N)$ theories
with spinors, and $G_2$ with five fundamentals~\cite{Pouliot,G2}.

We now present our arguments which enable us to identify other
theories which s-confine. Except for the discussion of
generalizations at the end of this letter we limit our attention to theories
with one gauge group and vanishing tree level superpotential.

The first argument follows from the requirement of smoothness of the
confining superpotential at the origin of moduli space. In the absence
of a tree level superpotential and with only one gauge group, the global
symmetries and holomorphy are sufficient to completely determine the form
of any non-perturbatively generated superpotential~\cite{ADS}. For a
theory with gauge group $G$ and matter fields $\phi_i$ this
superpotential is
\beq
  W \propto \left( \prod_i {\phi_i}^{\mu_i} \right)^{2 / ( \sum_j \mu_j -\mu(G) )} ,
\label{eq:ADS}
\eeq
where $\mu(G)$ is the Dynkin index\footnote{We normalize the index of
the fundamental representation to 1.}
of the adjoint representation
of $G$, and $\mu_i$ are the indices of the representations of the $\phi_i$.
Note that there may be several (or zero) possible contractions of
gauge indices, thus the superpotential can be a sum of several terms. 
We require the coefficient of this superpotential to be non-vanishing, then
holomorphy at the origin implies
that the exponents of all fields $\phi_i$ are positive integers.
Therefore\footnote{\samepage{Other solutions exist if all $\mu_i$ have
a common divisor $d$,
then for $\sum_j \mu_j -\mu(G)=d\ {\rm or}\ 2 d$ the superpotential
Eq.~\ref{mus} may be regular. We will argue at the end of the paper that these
solutions generically do not yield s-confining theories. Another possibility is that
the coefficient of the superpotential above vanishes. We will consider this
special case in our discussions at the end as well.}},
$\sum_j \mu_j -\mu(G)=1\ {\rm or}\ 2$, and for $SU$ and $Sp$ theories
anomaly cancellation further constrains
\beq
\sum_j \mu_j -\mu(G) = 2 .
\label{mus}
\eeq
This formula constitutes a necessary condition for s-confinement,
it enables us to rule out most theories immediately.
For example, for SQCD we find that the only candidate theory
is the theory with $F=N+1$.  
Unfortunately, Eq.~\ref{mus} is not a sufficient condition.
An example for a theory which satisfies Eq.~\ref{mus} but does not s-confine
is $SU(N)$ with an adjoint superfield and one flavor. This theory
is easily seen to be in an Abelian Coulomb phase for generic VEVs of the
adjoint scalars and vanishing VEVs for the fundamentals.

We could now simply examine all theories that satisfy Eq.~\ref{mus} by
finding all independent gauge invariants and checking if this ansatz for the
confining spectrum matches the anomalies. Apart from being very cumbersome,
this method is also not very useful to demonstrate that a given theory satisfying
Eq.~\ref{mus} is not s-confining.

A better strategy relies on our second observation. An s-confining theory with
a smooth description in terms of gauge invariants at the origin must also
be s-confining everywhere on its moduli space. This is because the
confining superpotential at the origin which is a simple polynomial in the fields
is analytical everywhere, and no additional massless states are present anywhere
on the moduli space.
Therefore, the theory restricted to a particular flat direction must have a
smooth description as well. This observation has two very useful applications.

First, if we have a theory that s-confines and we know its confined spectrum and
superpotential, we can easily find new s-confining theories by going
to different points on moduli space. In the ultraviolet description,
the gauge group is broken to a subgroup of the original group, some
matter fields are eaten by the Higgs mechanism, and the remaining ones
decompose under the unbroken subgroup. The corresponding confined
description is obtained by simply finding the corresponding point
on the moduli space of the confined theory. The global symmetries will be
broken in the same way, and some fields may be massive and can be
integrated out. This newly found confined theory is guaranteed to pass all
the standard consistency checks because they are a subset of the consistency
checks for the original theory. For example, the anomalies of the new
s-confining theory are guaranteed to match: the unbroken global symmetries
are a subgroup of original global symmetries, and the anomalies under the
subgroup are left unchanged -- both in the infrared and ultraviolet descriptions --         
because the fermions which obtain masses give cancelling
contributions to the anomalies.

Second, the above observation can be turned around to provide another
necessary condition for s-confinement. If anywhere on the moduli space
of a given theory we find a theory which is not s-confining or completely
higgsed, we know that the original theory cannot be
s-confining either.  

Let us study some examples. Suppose we knew that $SU(N)$ with $N+1$
flavors for some large $N$ is s-confining, then we could immediately
conclude that the theories with $n<N$ also s-confine.
We simply need to give a VEV to some of the quark-antiquark
pairs to break $SU(N)$ to any $SU(n)$ subgroup. The quarks with
vevs are eaten, leaving $n+1$ flavors and some singlets. 
We remove these singlets by adding ``mirror" superfields with opposite
global charges and giving them a mass. We now identify
the corresponding point on the moduli space of the
confined $SU(N)$ theory. Some fields obtain masses from the superpotential of
Eq.~\ref{SQCDpot} when we expand around the new point in moduli space. After
integrating the massive fields and removing the fields corresponding to the
singlets in the ultraviolet theory via masses with mirror partners,
we obtain the correct confined description of $SU(n)$.

A non-trivial example of a theory which can be shown to not s-confine is
$SU(4)$ with three antisymmetric tensors and two flavors. This theory
satisfies Eq.~\ref{mus} and is therefore a candidate for s-confinement. 
By giving a VEV to an antisymmetric tensor we can flow from this theory
to $Sp(4)$ with two antisymmetric tensors and four fundamentals. VEVs for the
other antisymmetric tensors let us flow further to $SU(2)$ with eight fundamentals
which is known to be at an interacting fixed point in the infrared.
We conclude that the $SU(4)$ and $Sp(4)$ theories and all theories that flow to them
cannot be s-confining either. This allows us to rule out the following chain of
theories, all of which are gauge anomaly free and satisfy Eq.~\ref{mus}
\beq
\begin{array}{ccccccccc}
              SU(7) & \to & SU(6) & \to & SU(5) & \to & SU(4) & \to & Sp(4) \\
        \Ythreea \, \, 2\, \Yfund \, \,  4\, \overline{\Yfund} & &
        \Ythreea  \, \, \Yasymm \, \,  \Yfund \, \, 3\, \overline{\Yfund} & &
        2\, \Yasymm  \, \, \overline{\Yasymm} \, \, \Yfund \, \, 2\, \overline{\Yfund} & &
        3\, \Yasymm \, \,  2\, \Yfund \, \, 2\, \overline{\Yfund} & &
        2\, \Yasymm \, \,  4\, \Yfund
            \end{array} 
\eeq
Note that a VEV for one of the quark flavors of the $SU(4)$ theory lets us
flow to
an $SU(3)$ theory with four flavors which is s-confining. We must therefore
be careful, when we find a flow to an s-confining theory, it does not follow that
the original theory is s-confining as well. The flow is only a necessary condition.
However, we suspect that a theory with a single gauge group and no tree-level
superpotential is s-confining if it is found to flow to s-confining theories in all directions
of its moduli space. We do not know of any counter examples.

Armed with formula in Eq.~\ref{mus} and our observation on flows of
s-confining theories, we were
able to find all s-confining $SU$ and $Sp$ gauge theories with a single gauge group and
no tree-level superpotential for arbitrary tensor representations. To achieve this, we first
found all possible matter contents satisfying Eq.~\ref{mus}. We list all these theories in Table 1.
We then studied the possible flows of these theories and discarded all those with flows to
theories which do not s-confine. This process eliminated all except about a dozen theories for
which we then explicitly determined the independent gauge invariants and matched anomalies
to find the confining spectra. These results are summarized in Table 1.

\begin{table}
\vspace*{-1cm}
\begin{tabular}{|ll|l|} \hline
$SU(N)$ & $(N+1) (\Yfund + \overline{\Yfund})$ & s-confining \\
$SU(N)$ & $\Yasymm + N\, \overline{\Yfund} + 4\, \Yfund $ & s-confining \\
$SU(N)$ & $\Yasymm + \overline{\Yasymm} + 3 (\Yfund + \overline{\Yfund})$ &
  s-confining \\
$SU(N)$ & Adj  $+\Yfund + \overline{\Yfund}$ & Coulomb branch \\ \hline
$SU(4)$ & Adj $+ \Yasymm $ & Coulomb branch \\
$SU(4)$ & $3\, \Yasymm + 2 (\Yfund + \overline{\Yfund})$ &
   $SU(2)$: $8\, \Yfund$ \\
$SU(4)$ & $ 4\, \Yasymm + \Yfund + \overline{\Yfund}$ &
   $SU(2)$: $\Ysymm + 4\, \Yfund$  \\
$SU(4)$ & $ 5\, \Yasymm $ & Coulomb branch \\
$SU(5)$ & $ 3 (\Yasymm + \overline{\Yfund}) $ & s-confining \\
$SU(5)$ & $ 2\, \Yasymm + 2\, \Yfund + 4\, \overline{\Yfund}$ & s-confining \\
$SU(5)$ & $ 2 (\Yasymm + \overline{\Yasymm})$ & 
   $Sp(4)$: $3\, \Yasymm + 2\, \Yfund$ \\
$SU(5)$ & $2\, \Yasymm + \overline{\Yasymm} + 2\, \overline{\Yfund} +
  \Yfund$ &  $SU(4)$: $3\, \Yasymm + 
  2 (\Yfund + \overline{\Yfund})$ \\
$SU(6)$ & $2\, \Yasymm + 5\, \overline{\Yfund} + \Yfund$ &
  s-confining \\
$SU(6)$ & $ 2\, \Yasymm + \overline{\Yasymm} + 2\, \overline{\Yfund}$ &
   $SU(4)$: $3\, \Yasymm + 2 (\Yfund + \overline{\Yfund})$ \\
$SU(6)$ & $\Ythreea + 4 (\Yfund + \overline{\Yfund})$ & s-confining \\
$SU(6)$ & $\Ythreea + \Yasymm +  3\, \overline{\Yfund} + \Yfund$
  &  $SU(5)$: $2\, \Yasymm +
  \overline{\Yasymm} + 2\, \overline{\Yfund} + \Yfund $ \\
$SU(6)$ & $\Ythreea + \Yasymm + \overline{\Yasymm}$ & 
  $Sp(6)$: $\Ythreea + \Yasymm + \Yfund$ \\
$SU(6)$ & $2\, \Ythreea + \Yfund + \overline{\Yfund}$ &  $SU(5)$:
  $  2 (\Yasymm + \overline{\Yasymm})$ \\
$SU(7)$ & $2 (\Yasymm + 3\, \overline{\Yfund})$ & s-confining \\
$SU(7)$ & $ \Ythreea + 4\, \overline{\Yfund} + 2\, \Yfund$ &
   $SU(6)$: $\Ythreea + \Yasymm + 3\, \overline{\Yfund}+ \Yfund$ \\
$SU(7)$ & $\Ythreea + \overline{\Yasymm} + \Yfund$ & $Sp(6)$: $\Ythreea + 
   \Yasymm + \Yfund$ \\
  \hline \hline
$Sp(2N)$ & $(2N+4)\, \Yfund$ & s-confining \\
$Sp(2N)$ & $\Yasymm +6\, \Yfund $ & s-confining \\
$Sp(2N)$ & $\Ysymm +2\, \Yfund $ & Coulomb branch \\ \hline 
$Sp(4)$ & $2\, \Yasymm +4\, \Yfund $& $SU(2)$: $8\, \Yfund$ \\
$Sp(4)$ & $3\, \Yasymm +2\, \Yfund$& $SU(2)$: $\Ysymm +4\, \Yfund$ \\
$Sp(4)$ & $ 4\, \Yasymm$ & $SU(2)$: $2\, \Ysymm$ \\
$Sp(6)$ & $2\, \Yasymm +2\,\Yfund$ & $Sp(4)$: $2\, \Yasymm +4\, \Yfund $ \\
$Sp(6)$ & $\Ythreea +5\,\Yfund$ & $Sp(4)$: $2\, \Yasymm +4\, \Yfund $ \\
$Sp(6)$ & $\Ythreea +\Yasymm +\Yfund$ & $SU(2)$: $\Ysymm +4\, \Yfund$ \\
$Sp(6)$ & $2\, \Ythreea$ & $SU(3)$: $\Ysymm + \overline{\Ysymm}$ \\
$Sp(8)$ & $2\, \Yasymm $& $Sp(4)$: $ 5\,\Yasymm$\\ 
\hline
\end{tabular}
\caption{All $SU$ and $Sp$ theories satisfying $\sum_j \mu_j -\mu(G) = 2$.
Note that this list is finite because the indices of higher index tensor
representations grow very rapidly with the size of the gauge group.
We list the gauge group and the field content of the theories in the first column.
In the second column, we indicate which theories are s-confining.
For the remaining ones we give the flows to non-confining theories
or indicate that there is a Coulomb branch on the moduli space.}
\end{table}

Six of the ten theories which s-confine are new%
\footnote{Some of these results were obtained independently by
A. Nelson~\cite{privcomm}.}: $SU(N)$ with
$\Yasymm +  \overline{\Yasymm}\ + 3\,\Yfund +  3\, \overline{\Yfund}$,
$SU(7)$ with $2\, \Yasymm +  6\, \overline{\Yfund}$, $SU(6)$ with
$2\, \Yasymm +  \Yfund + 5\, \overline{\Yfund}$,
$SU(6)$ with $\Ythreea +  4\, \Yfund +  4\, \overline{\Yfund}$, $SU(5)$ with
$2\, \Yasymm + 2\, \Yfund +  4\, \overline{\Yfund}$, and
$SU(5)$ with $3\, \Yasymm +  3\, \overline{\Yfund}$.
For the theories which do not s-confine we indicated the method by which we
obtained this result: either by noting that the theory has a branch with
only unbroken $U(1)$ gauge groups, or else by flowing along a flat direction
to a theory with smaller non-Abelian gauge group which does not s-confine.

Detailed results on the new theories including the confining spectra, superpotentials,
various flows, and consistency checks will be reported elsewhere \cite{toappear}. Here, we
just point out a few salient features.

Most of the new s-confining theories contain vector-like matter. Perturbing
these theories by adding mass terms for some of the vector-like matter,
we easily obtain exact results on the theories with the matter integrated out.
Among the theories that we find in this way are new
theories which confine with chiral symmetry breaking, theories with runaway vacua,
and theories which confine without chiral symmetry breaking and vanishing
superpotentials. Since many of the new theories presented here are chiral, they
can be used to find models of dynamical supersymmetry breaking along the lines
of Refs.~\cite{dsb}. Examples for such supersymmetry breaking theories will also
be included in the detailed paper~\cite{toappear}. 
Our s-confining theories might be used for building extensions of the standard
model with composite quarks and leptons~\cite{composite}.

Finally, we comment on possible exceptions and generalizations of our arguments.
A possible exception to our condition in Eq.~\ref{mus} arises, when all $\mu_i$ and
$\mu(G)$ have a common divisor. Then the superpotential Eq.~\ref{eq:ADS} can be
holomorphic even when $\sum_j \mu_j - \mu (G) \not= 2$. However, whereas Eq.~\ref{mus}
is preserved under most flows, the property that all $\mu$'s have a common divisor
is not. Therefore, such theories flow to theories which are not s-confining, and by
our second necessary condition the original theory is not s-confining either.

Another possibility is that the confining superpotential vanishes, and the confined
degrees of freedom are free in the infrared.
This can only happen if there are no classical constraints among the basic gauge invariant
operators which satisfy the 't~Hooft anomaly matching conditions, otherwise the
quantum solution would not have the correct classical limit. Examples
of theories which are believed to confine in this way can be found in the
literature~\cite{SO,ISS,toappear}.

Generalizations to $SO(N)$ groups are not completely straightforward because in
the case of $SO(N)$ theories ``exotic composites'' containing the chiral superfield
$W_{\alpha}$  might appear in the infrared spectrum and superpotential, thus modifying
our argument and result of Eq.~\ref{mus}.

Generalizations to theories with more than one gauge group or tree level superpotentials
are more difficult. The additional interactions break some of the global symmetries
which are now not sufficient to completely determine the functional form of the
confining superpotential. Another complication is that in these theories the flat
directions of the quantum theory are sometimes difficult to identify. Since our
second argument only applies to flows in directions which are on the quantum
moduli space, incorrect conclusions would be obtained from flows
along classical flat directions which are not flat in the quantum theory.

In summary, we have discussed general criteria for s-confinement and used them
to find all s-confining theories with $SU(N)$ or $Sp(2N)$ gauge groups. 

It is a pleasure to thank P. Cho, A. Cohen, N. Evans, L. Randall,
and R. Sundrum for useful discussions. We also thank B. Dobrescu,
A. Nelson, and J. Terning for comments on the manuscript.
C.C. and W.S. are supported in part by the U.S. Department of Energy
under cooperative
agreement \#DE-FC02-94ER40818. M.S. is supported by the U.S.
Department of Energy under grant \#DE-FG02-91ER40676.

\end{document}